%% file: paper.tex
\documentstyle [art11,epsf]{article}
\title	{$SU(N_c)$ gauge theories for all $N_c$.}
\author	{
\\
M. Teper\thanks{presented at Lattice 96}
\\
{\small\sl Theoretical Physics, University of Oxford,
	1 Keble Road, Oxford, OX1 3NP, U.K.}}
\date{}
\begin	{document}
\maketitle
\vskip 0.7in
\begin	{abstract}
\noindent
We show that $SU(N_c)$ gauge theories in 2+1 dimensions
are close to $N_c=\infty$ for $N_c\geq 2$. The dimensionful
coupling, $g^2$, is proportional to $1/N_c$, 
at large $N_c$, confirming
the usual diagram-based expectation. Preliminary 
calculations in 3+1 dimensions indicate that the
same is true there. 
\end	{abstract}

\vfill
Oxford Preprint Number: {\em OUTP--96--49P}
\hfill
hep-lat/9701004

\newpage
\section	{Introduction.}
\label		{intro}

The idea of considering $SU(N_c)$ gauge  
theories as perturbations in powers of $1/N_c$ 
around $N_c=\infty$ is an old one

\cite{tHo}.
It has been buoyed by the fact that, if one assumes 
confinement for all $N_c$, then the phenomenology
of the $SU(\infty)$ quark-gluon theory is strikingly 
similar to that of (the non-baryonic sector of) QCD
\cite{tHo,Wit}.
This makes it conceivable that the physically
interesting $SU(3)$ theory could be largely
understood by solving the much simpler
$SU(\infty)$ theory (for a review see
\cite{Cole}).
The fact that the lattice $SU(\infty)$ theory
can be re-expressed as a single plaquette
theory
\cite{EK},
provided the basis of a number of interesting
computational explorations of that theory
(for a review see
\cite{Das}).
Unfortunately this latter scheme is limited by 
the fact that it does not tell us anything 
about the size of the corrections to the
$N_c=\infty$ limit.

Over the last few years I have investigated
$D=2+1$ $SU(N_c)$ gauge theories by Monte
Carlo simulation. These theories are
similar to the corresponding $D=3+1$ theories:
the interaction strength vanishes at short
distances, while at long distances there
appears to be a non-perturbative linear
confining potential. The coupling sets the
overall mass scale in both cases: in $D=2+1$ 
$g^2$ has dimensions of mass while in $D=3+1$
the scale invariance is anomalous, so the
coupling runs and the rate at which it runs 
introduces a mass scale (i.e. the $\Lambda$
parameter). On the other hand the computational 
requirements are obviously much less
in $D=2+1$ than in $D=3+1$. This has enabled me to
obtain a detailed continuum mass spectrum
that is much more accurate than that available 
in 4 dimensions.

The $C=+$ sector of the light mass spectrum turned 
out be quite similar for both $SU(2)$ 
\cite{MT2G}
and $SU(3)$
\cite{MT3G}.
%.
(This also appears to be the case in $D=3+1$, 
although the comparison there is weakened by 
the much larger errors.) If the reason for this is

that both are close to the $N_c=\infty$ limit, then
this provides an economical understanding of the spectra
of $SU(N_c)$ gauge theories for all $N_c$, i.e.
there is a common spectrum with small corrections.

At the same time, many theoretical approaches are simpler
in the large $N_c$ limit. For example, there has been
recent progress in calculating the
large $N_c$ mass spectrum using light front quantisation
techniques
\cite{Dalley}.

Motivated by all this I have investigated the $N_c\to\infty$
limit in the most direct fashion: by performing 
some $SU(4)$ calculations. In $D=2+1$ I have found that the $SU(3)$
and $SU(4)$ spectra are almost identical in both
$C=+$ and $C=-$ sectors. As we shall see in the next section,
we can describe the variations between the $N_c=2,3$ and 4
mass spectra by a simple $O(1/N^2_c)$ correction. This
is the expected form for the leading large-$N_c$ correction
in pure gauge theories
\cite{Cole}.
In the final section I present some  preliminary results
for the $SU(4)$ theory in 4 dimensions. The light masses
considered, and the topological susceptibility, indicate
that here too there is a rapid approach to the large-$N_c$
limit.

\section	{2+1 dimensions.}

\label	{3dim}

Consider the string tension. Using smeared Polyakov loops
in the standard fashion
\cite{CMT},
I have calculated the continuum string tension, $\sigma$,
in units of $g^2$ for $SU(2)$
\cite{MT2K},
, $SU(3)$ and $SU(4)$
\cite{MT3G},
gauge theories in $D=2+1$. The values of $\surd\sigma/g^2$
are plotted in 
Fig.~\ref{fig_plot_string3}
against the
number of colours, $N_c$. We see that there is an
approximate linear rise with $N_c$. We obtain a
good fit with
$$ {{\surd \sigma} \over g^2} = 0.1974(12) N_c 
-{0.120(8) \over N_c}. \eqno(1) $$
We would obtain a similar behaviour with other light
glueball masses: we focus on the string tension
since it is the most accurate quantity in our
calculations.

Some observations.

$\bullet$ For large $N_c$, $\surd\sigma \propto g^2 N_c$.
That is to say, the overall mass scale of the theory, call
it $\mu$, is proportional to $g^2 N_c$. So in units of the
mass scale of the theory,
$$ g^2 \propto {\mu \over N_c}. \eqno(2)$$
This is the usual diagram-based expectation, but here the
argument is entirely non-perturbative.

$\bullet$ Our correction term has the theoretically-expected 
functional form, i.e. it is $O(1/N^2_c)$ relative to the
leading term. If we try a fit with a $O(1/N_c)$
correction instead, we obtain an unacceptably poor $\chi^2$.
(But higher powers are not necessarily excluded by our `data'.)
These expectations apply not only to the continuum limit,
but also to lattice corrections, and our lattice
values do indeed show that the leading lattice correction
is $O(ag^2N_c)$.

$\bullet$ The coefficients of the leading term and the
first correction are comparable, suggesting a well-behaved
expansion in powers of $1/N_c$.

We have also fitted the lightest glueball masses,
with some cases shown in 
Fig.~\ref{fig_plot_glue3}.
For example we find for the lightest glueball
$$ {{m_{0^{++}}}\over{\surd\sigma}}
= 4.046(70) + {{2.67(28)} \over N^2_c} \eqno(4)$$
and for its first excitation
$$ {{{m^\ast}_{0^{++}}}\over{\surd\sigma}}
= 6.28(16)  + {{2.16(86)} \over N^2_c}. \eqno(5)$$
This provides us with mass predictions not just for
$N_c=\infty$ but for $all$ $N_c$.

\section	{3+1 dimensions.}

\label	{4dim}

What we know about 4 dimensional gauge theories  
is currently much less precise. As far as continuum 
properties are concerned, reasonably accurate 
quantities include the string tension,
the lightest scalar and tensor glueballs and the
deconfining transition. As in the $D=2+1$ case,
the $SU(2)$ and $SU(3)$ values are within $\sim 20\%$ of each 
other. This encourages an investigation of the
$SU(4)$ theory. 

In $D=3+1$, SU(4) calculations are slow and we
have only performed a very preliminary study.

We use the standard plaquette action, and so our
first potential hurdle is the presence of the well-known 
bulk transition that occurs 
as we increase $\beta$ from the strong coupling regime
towards weak coupling. We have performed a scan
on a $10^4$ lattice which locates this 
transition at $\beta_c = 10.4 \pm 0.1$. This
turns out to be a rather strong coupling and so
it does not complicate our subsequent calculations.

We performed 4000,6000 and 3000 sweeps respectively 
on $10^4$,$12^4$ and $16^4$ 
lattices at $\beta=$10.7,10.9 and 11.1.
We carried out correlator measurements every 5'th sweep.

$\bullet$ We obtain string tensions
$a\surd\sigma =$ 0.296(14), 0.229(7) and 0.196(7)
at $\beta =$10.7, 10.9 and 11.1 respectively.
(Here $a$ is the lattice spacing.) Comparing
with $SU(2)$ and $SU(3)$
\cite{CMT},
and using $\beta = 2N_c/g^2$, we see that the
bare coupling, $g^2(a)$, for the same lattice
spacing (in physical units) varies roughly
as $g^2(a) \propto 1/N_c$. A better 
coupling to use is the mean-field improved one
\cite{MFI},
$g^2_I(a)$, obtained from $g^2(a)$ by dividing
it by $<{1\over N_c}TrU_p>$. For example $a\surd\sigma$
is approximately the same at $\beta=2.47,6.0,11.0$
for $N_c =2,3,4$ respectively. This corresponds
to $g^2_I(a) \simeq 2.509, 1.684, 1.244$ respectively.
These vary almost $exactly$ as $1/N_c$, confirming
the usual diagrammatic expectation.
 
\begin	{figure}[p]
\begin	{center}
\leavevmode
\input	{plot_glue4}
\end	{center}
\caption{Lightest scalar ($\bullet$) and tensor ($\circ$)
 glueball masses in $D=4$. Continuum values for $N_c=2,3$
 and lattice values ($\beta=10.9$ and $\beta=11.1$)
 for $N_c=4$.}
\label	{fig_plot_glue4}
\end 	{figure}
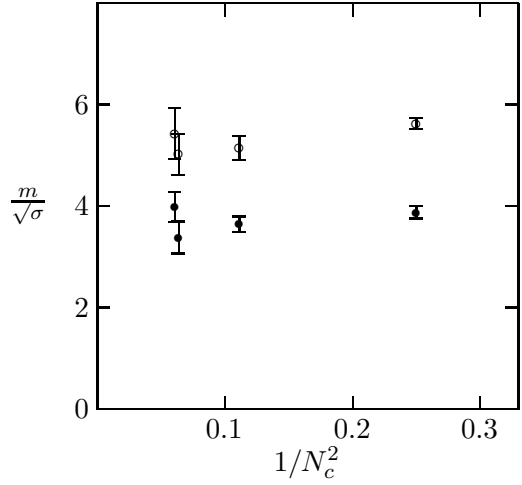

$\bullet$ In
Fig.~\ref{fig_plot_glue4}
I plot the ratios of scalar and tensor glueball
masses to $\surd\sigma$. The $N_c=2,3$ values
are continuum extrapolations, while the two 
$N_c=4$ values are simply those obtained at 
$\beta=10.9, 11.1$ since our calculations
are not accurate enough for a continuum
extrapolation. Although the $N_c=4$ errors are
large, we are certainly consistent with
the $N_c$ variation being given by a simple
$1/N_c^2$ correction. 

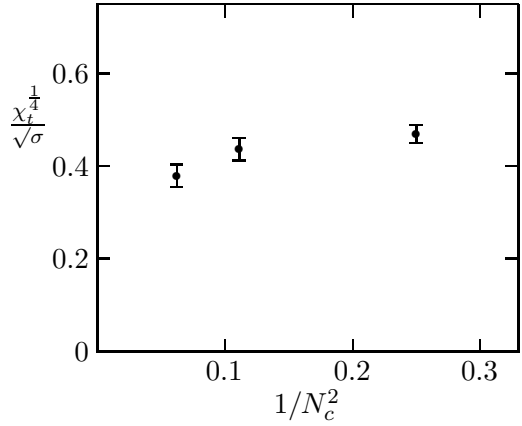
\begin	{figure}[p]
\begin	{center}
\leavevmode
\input	{plot_top4}
\end	{center}
\caption{The topological susceptibility: continuum values
for $N_c=2,3$ and averaged lattice value for $N_c=4$.}
\label	{fig_plot_top4}
\end 	{figure}

$\bullet$ We have also calculated the topological
susceptibility, $\chi_t \equiv <Q^2>/vol$, using the 
standard cooling method. In
Fig.~\ref{fig_plot_top4}
I show the dimensionless ratio $\chi_t^{1/4}/\surd\sigma$
for $\beta=10.9$ and 11.1, and compare these values
to the $N_c=2,3$ continuum values. (Warning: for such a 
short Monte Carlo run the errors are certain to be
underestimated.) Again there is
consistency  with a simple $1/N_c^2$ correction.
We remark that for $SU(4)$ one expects, semiclassically, very 
few small instantons and this is confirmed in our cooling
calculations. This has the advantage that the 
lattice ambiguities that arise when instantons 
are not much larger than $a$
are reduced compared to $SU(3)$ and dramatically reduced
as compared to $SU(2)$. The large-$N_c$ physics of
topology (and the related meson physics) is of obvious
interest.

\section	{Conclusions.}

\label		{conc}

We have calculated the mass spectra of gauge theories
with $N_c=2,3,4$ in 3 dimensions. We have found that
the string tension and lightest masses are accurately
given by a $O(1/N_c^2)$ correction to the $N_c=\infty$
limit. This provides a unified understanding of all
these theories in terms of just one theory with
modest corrections to it. In practical terms
it means that we know the corresponding
masses for $all$ values of $N_c$.

Preliminary calculations in 4 dimensions
suggest that the situation is the same in that case. 
Here we have investigated not only the lightest masses
but the topological susceptibility as well. We remark
that the lattice calculation of $Q$ becomes rapidly cleaner
as $N_c$ increases.

Our work provides explicit confirmation of 
the old, bold conjecture
\cite{Wit}
that, in the context of the $1/N_c$ expansion, 3 is
a large number. Moreover, we provide a significant 
extension of that conjecture: 2 is also a large number.

%
%
%
%
%\begin{figure}[p]
%\begin{center}
%\leavevmode
%\epsfxsize=2.5in
%\epsffile{mon_b24_l12_max.ps}
%\epsfxsize=2.5in
%\epsffile{mon_b24_l12_diff.ps}
%\end{center}
%%
%\caption{Comparison of the loop spectra in $D=4$ on a $12^4$ lattice
%at $\beta_4 = 2.4$. Plotted are the spectra for $\{ R_{max} \}$
%with $N_{GT} = 5$, and the difference gas between the two extremes
%also with $N_{GT} = 5$}
%\label{diff_spect_4}
%\end{figure}
%
\begin	{figure}[p]
\begin	{center}
\leavevmode
\input	{plot_string3}
\end	{center}
\caption{Continuum string tension versus number of colours
in $D=2+1$. Line is the fit in eqn(1).}
\label	{fig_plot_string3}
\end 	{figure}
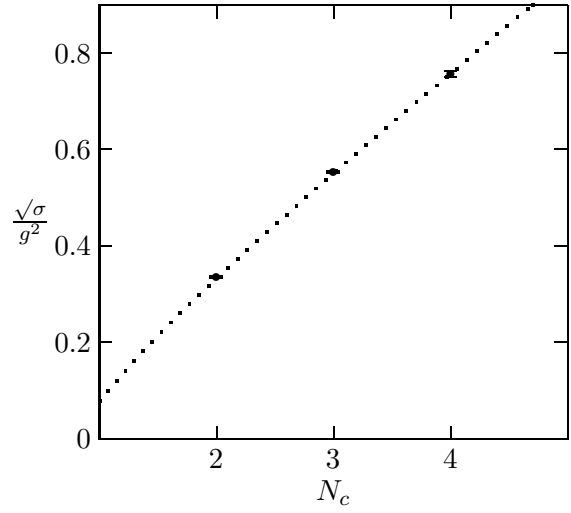

\begin	{figure}[p]
\begin	{center}
\leavevmode
\input	{plot_glue3}
\end	{center}
\caption{Some continuum glueball masses, in $D=3$, for 2,3,4
colours: $0^{++}$($\bullet$), $0^{++*}$($\times$), $2^{++}$($\star$), 
$0^{-+}$($\diamond$) and linear fits. }

\label	{fig_plot_glue3}
\end 	{figure}
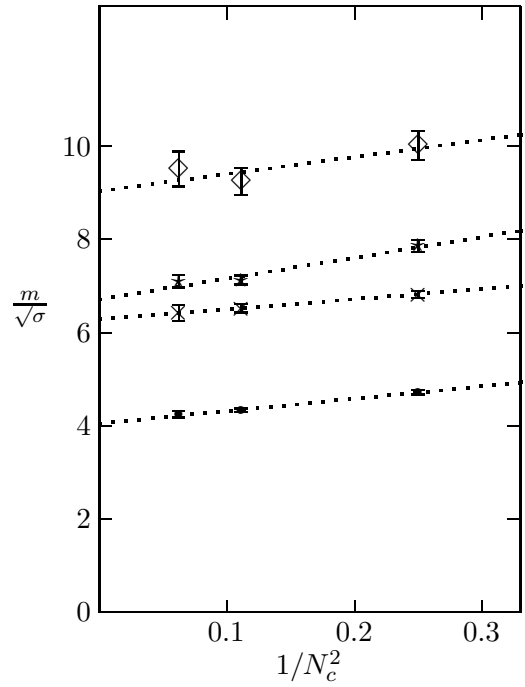

\end{document}

%% file: plot_string3.tex
% GNUPLOT: LaTeX picture
\setlength{\unitlength}{0.240900pt}
\ifx\plotpoint\undefined\newsavebox{\plotpoint}\fi
\sbox{\plotpoint}{\rule[-0.200pt]{0.400pt}{0.400pt}}%
\begin{picture}(900,720)(0,0)
\font\gnuplot=cmr10 at 12pt
\gnuplot
\sbox{\plotpoint}{\rule[-0.200pt]{0.400pt}{0.400pt}}%
\put(120.0,31.0){\rule[-0.200pt]{4.818pt}{0.400pt}}
\put(108,31){\makebox(0,0)[r]{{$0$}}}
\put(836.0,31.0){\rule[-0.200pt]{4.818pt}{0.400pt}}
\put(120.0,183.0){\rule[-0.200pt]{4.818pt}{0.400pt}}
\put(108,183){\makebox(0,0)[r]{{$0.2$}}}
\put(836.0,183.0){\rule[-0.200pt]{4.818pt}{0.400pt}}
\put(120.0,334.0){\rule[-0.200pt]{4.818pt}{0.400pt}}
\put(108,334){\makebox(0,0)[r]{{$0.4$}}}
\put(836.0,334.0){\rule[-0.200pt]{4.818pt}{0.400pt}}
\put(120.0,486.0){\rule[-0.200pt]{4.818pt}{0.400pt}}
\put(108,486){\makebox(0,0)[r]{{$0.6$}}}
\put(836.0,486.0){\rule[-0.200pt]{4.818pt}{0.400pt}}
\put(120.0,637.0){\rule[-0.200pt]{4.818pt}{0.400pt}}
\put(108,637){\makebox(0,0)[r]{{$0.8$}}}
\put(836.0,637.0){\rule[-0.200pt]{4.818pt}{0.400pt}}
\put(304.0,31.0){\rule[-0.200pt]{0.400pt}{4.818pt}}
\put(304,19){\makebox(0,0){\shortstack{\\ \\ \\ {$2$}}}}
\put(304.0,693.0){\rule[-0.200pt]{0.400pt}{4.818pt}}
\put(488.0,31.0){\rule[-0.200pt]{0.400pt}{4.818pt}}
\put(488,19){\makebox(0,0){\shortstack{\\ \\ \\ {$3$}}}}
\put(488.0,693.0){\rule[-0.200pt]{0.400pt}{4.818pt}}
\put(672.0,31.0){\rule[-0.200pt]{0.400pt}{4.818pt}}
\put(672,19){\makebox(0,0){\shortstack{\\ \\ \\ {$4$}}}}
\put(672.0,693.0){\rule[-0.200pt]{0.400pt}{4.818pt}}
\put(120.0,31.0){\rule[-0.200pt]{177.302pt}{0.400pt}}
\put(856.0,31.0){\rule[-0.200pt]{0.400pt}{164.294pt}}
\put(120.0,713.0){\rule[-0.200pt]{177.302pt}{0.400pt}}
\put(12,372){\makebox(0,0){{${{\surd\sigma} \over g^2}$}}}
\put(488,-53){\makebox(0,0){{$N_c$}}}
\put(120.0,31.0){\rule[-0.200pt]{0.400pt}{164.294pt}}
\put(304,285){\circle*{12}}
\put(488,450){\circle*{12}}
\put(672,604){\circle*{12}}
\put(304.0,284.0){\rule[-0.200pt]{0.400pt}{0.482pt}}
\put(294.0,284.0){\rule[-0.200pt]{4.818pt}{0.400pt}}
\put(294.0,286.0){\rule[-0.200pt]{4.818pt}{0.400pt}}
\put(488.0,449.0){\rule[-0.200pt]{0.400pt}{0.723pt}}
\put(478.0,449.0){\rule[-0.200pt]{4.818pt}{0.400pt}}
\put(478.0,452.0){\rule[-0.200pt]{4.818pt}{0.400pt}}
\put(672.0,599.0){\rule[-0.200pt]{0.400pt}{2.409pt}}
\put(662.0,599.0){\rule[-0.200pt]{4.818pt}{0.400pt}}
\put(662.0,609.0){\rule[-0.200pt]{4.818pt}{0.400pt}}
\sbox{\plotpoint}{\rule[-0.500pt]{1.000pt}{1.000pt}}%
\put(120,90){\usebox{\plotpoint}}
\put(120.00,90.00){\usebox{\plotpoint}}
\put(132.84,106.30){\usebox{\plotpoint}}
\multiput(135,109)(12.743,16.383){0}{\usebox{\plotpoint}}
\put(146.17,122.17){\usebox{\plotpoint}}
\multiput(150,126)(12.743,16.383){0}{\usebox{\plotpoint}}
\put(159.62,137.95){\usebox{\plotpoint}}
\multiput(165,144)(13.668,15.620){0}{\usebox{\plotpoint}}
\put(173.33,153.52){\usebox{\plotpoint}}
\multiput(179,160)(14.676,14.676){0}{\usebox{\plotpoint}}
\put(187.55,168.63){\usebox{\plotpoint}}
\put(201.75,183.75){\usebox{\plotpoint}}
\multiput(202,184)(13.668,15.620){0}{\usebox{\plotpoint}}
\put(215.91,198.91){\usebox{\plotpoint}}
\multiput(217,200)(14.676,14.676){0}{\usebox{\plotpoint}}
\put(230.59,213.59){\usebox{\plotpoint}}
\multiput(232,215)(14.676,14.676){0}{\usebox{\plotpoint}}
\put(245.26,228.26){\usebox{\plotpoint}}
\multiput(246,229)(14.676,14.676){0}{\usebox{\plotpoint}}
\put(259.94,242.94){\usebox{\plotpoint}}
\multiput(261,244)(15.620,13.668){0}{\usebox{\plotpoint}}
\put(275.10,257.10){\usebox{\plotpoint}}
\multiput(276,258)(15.620,13.668){0}{\usebox{\plotpoint}}
\put(290.26,271.26){\usebox{\plotpoint}}
\multiput(291,272)(14.676,14.676){0}{\usebox{\plotpoint}}
\put(305.38,285.46){\usebox{\plotpoint}}
\multiput(306,286)(14.676,14.676){0}{\usebox{\plotpoint}}
\put(320.55,299.61){\usebox{\plotpoint}}
\multiput(321,300)(14.676,14.676){0}{\usebox{\plotpoint}}
\put(335.72,313.76){\usebox{\plotpoint}}
\multiput(336,314)(14.676,14.676){0}{\usebox{\plotpoint}}
\multiput(343,321)(14.676,14.676){0}{\usebox{\plotpoint}}
\put(350.47,328.35){\usebox{\plotpoint}}
\multiput(358,334)(14.676,14.676){0}{\usebox{\plotpoint}}
\put(366.09,341.95){\usebox{\plotpoint}}
\multiput(373,348)(15.759,13.508){0}{\usebox{\plotpoint}}
\put(381.77,355.55){\usebox{\plotpoint}}
\multiput(388,361)(14.676,14.676){0}{\usebox{\plotpoint}}
\put(397.06,369.54){\usebox{\plotpoint}}
\multiput(403,374)(14.676,14.676){0}{\usebox{\plotpoint}}
\put(412.43,383.43){\usebox{\plotpoint}}
\multiput(417,388)(16.604,12.453){0}{\usebox{\plotpoint}}
\put(428.03,397.03){\usebox{\plotpoint}}
\multiput(432,401)(16.604,12.453){0}{\usebox{\plotpoint}}
\put(443.64,410.64){\usebox{\plotpoint}}
\multiput(447,414)(16.604,12.453){0}{\usebox{\plotpoint}}
\put(459.24,424.24){\usebox{\plotpoint}}
\multiput(462,427)(15.759,13.508){0}{\usebox{\plotpoint}}
\put(474.75,438.03){\usebox{\plotpoint}}
\multiput(477,440)(15.759,13.508){0}{\usebox{\plotpoint}}
\put(490.43,451.62){\usebox{\plotpoint}}
\multiput(492,453)(15.759,13.508){0}{\usebox{\plotpoint}}
\put(506.11,465.22){\usebox{\plotpoint}}
\multiput(507,466)(15.759,13.508){0}{\usebox{\plotpoint}}
\multiput(514,472)(15.759,13.508){0}{\usebox{\plotpoint}}
\put(521.85,478.75){\usebox{\plotpoint}}
\multiput(529,485)(15.759,13.508){0}{\usebox{\plotpoint}}
\put(537.53,492.34){\usebox{\plotpoint}}
\multiput(544,498)(15.759,13.508){0}{\usebox{\plotpoint}}
\put(553.36,505.77){\usebox{\plotpoint}}
\multiput(559,510)(14.676,14.676){0}{\usebox{\plotpoint}}
\put(568.89,519.47){\usebox{\plotpoint}}
\multiput(573,523)(16.604,12.453){0}{\usebox{\plotpoint}}
\put(584.77,532.77){\usebox{\plotpoint}}
\multiput(588,536)(16.604,12.453){0}{\usebox{\plotpoint}}
\put(600.70,546.03){\usebox{\plotpoint}}
\multiput(603,548)(15.620,13.668){0}{\usebox{\plotpoint}}
\put(616.39,559.62){\usebox{\plotpoint}}
\multiput(618,561)(16.604,12.453){0}{\usebox{\plotpoint}}
\put(632.11,573.11){\usebox{\plotpoint}}
\multiput(633,574)(15.759,13.508){0}{\usebox{\plotpoint}}
\multiput(640,580)(16.604,12.453){0}{\usebox{\plotpoint}}
\put(648.21,586.18){\usebox{\plotpoint}}
\multiput(655,592)(15.620,13.668){0}{\usebox{\plotpoint}}
\put(663.89,599.77){\usebox{\plotpoint}}
\multiput(670,605)(16.604,12.453){0}{\usebox{\plotpoint}}
\put(679.92,612.92){\usebox{\plotpoint}}
\multiput(685,618)(15.759,13.508){0}{\usebox{\plotpoint}}
\put(695.48,626.61){\usebox{\plotpoint}}
\multiput(700,630)(15.759,13.508){0}{\usebox{\plotpoint}}
\put(711.43,639.88){\usebox{\plotpoint}}
\multiput(715,643)(15.759,13.508){0}{\usebox{\plotpoint}}
\put(727.43,653.07){\usebox{\plotpoint}}
\multiput(730,655)(15.759,13.508){0}{\usebox{\plotpoint}}
\put(742.89,666.89){\usebox{\plotpoint}}
\multiput(744,668)(16.604,12.453){0}{\usebox{\plotpoint}}
\put(758.97,679.98){\usebox{\plotpoint}}
\multiput(759,680)(16.604,12.453){0}{\usebox{\plotpoint}}
\multiput(767,686)(15.759,13.508){0}{\usebox{\plotpoint}}
\put(775.13,692.99){\usebox{\plotpoint}}
\multiput(782,699)(15.759,13.508){0}{\usebox{\plotpoint}}
\put(790.92,706.44){\usebox{\plotpoint}}
\multiput(797,711)(14.676,14.676){0}{\usebox{\plotpoint}}
\put(799,713){\usebox{\plotpoint}}
\end{picture}

%% file: plot_glue3.tex
% GNUPLOT: LaTeX picture
\setlength{\unitlength}{0.240900pt}
\ifx\plotpoint\undefined\newsavebox{\plotpoint}\fi
\sbox{\plotpoint}{\rule[-0.200pt]{0.400pt}{0.400pt}}%
\begin{picture}(825,990)(0,0)
\font\gnuplot=cmr10 at 12pt
\gnuplot
\sbox{\plotpoint}{\rule[-0.200pt]{0.400pt}{0.400pt}}%
\put(120.0,31.0){\rule[-0.200pt]{4.818pt}{0.400pt}}
\put(108,31){\makebox(0,0)[r]{{$0$}}}
\put(761.0,31.0){\rule[-0.200pt]{4.818pt}{0.400pt}}
\put(120.0,177.0){\rule[-0.200pt]{4.818pt}{0.400pt}}
\put(108,177){\makebox(0,0)[r]{{$2$}}}
\put(761.0,177.0){\rule[-0.200pt]{4.818pt}{0.400pt}}
\put(120.0,324.0){\rule[-0.200pt]{4.818pt}{0.400pt}}
\put(108,324){\makebox(0,0)[r]{{$4$}}}
\put(761.0,324.0){\rule[-0.200pt]{4.818pt}{0.400pt}}
\put(120.0,470.0){\rule[-0.200pt]{4.818pt}{0.400pt}}
\put(108,470){\makebox(0,0)[r]{{$6$}}}
\put(761.0,470.0){\rule[-0.200pt]{4.818pt}{0.400pt}}
\put(120.0,617.0){\rule[-0.200pt]{4.818pt}{0.400pt}}
\put(108,617){\makebox(0,0)[r]{{$8$}}}
\put(761.0,617.0){\rule[-0.200pt]{4.818pt}{0.400pt}}
\put(120.0,763.0){\rule[-0.200pt]{4.818pt}{0.400pt}}
\put(108,763){\makebox(0,0)[r]{{$10$}}}
\put(761.0,763.0){\rule[-0.200pt]{4.818pt}{0.400pt}}
\put(320.0,31.0){\rule[-0.200pt]{0.400pt}{4.818pt}}
\put(320,19){\makebox(0,0){\shortstack{\\ \\ \\ {$0.1$}}}}
\put(320.0,963.0){\rule[-0.200pt]{0.400pt}{4.818pt}}
\put(521.0,31.0){\rule[-0.200pt]{0.400pt}{4.818pt}}
\put(521,19){\makebox(0,0){\shortstack{\\ \\ \\ {$0.2$}}}}
\put(521.0,963.0){\rule[-0.200pt]{0.400pt}{4.818pt}}
\put(721.0,31.0){\rule[-0.200pt]{0.400pt}{4.818pt}}
\put(721,19){\makebox(0,0){\shortstack{\\ \\ \\ {$0.3$}}}}
\put(721.0,963.0){\rule[-0.200pt]{0.400pt}{4.818pt}}
\put(120.0,31.0){\rule[-0.200pt]{159.235pt}{0.400pt}}
\put(781.0,31.0){\rule[-0.200pt]{0.400pt}{229.337pt}}
\put(120.0,983.0){\rule[-0.200pt]{159.235pt}{0.400pt}}
\put(12,507){\makebox(0,0){{${m \over {\surd\sigma}}$}}}
\put(450,-53){\makebox(0,0){{$1/N^2_c$}}}
\put(120.0,31.0){\rule[-0.200pt]{0.400pt}{229.337pt}}
\put(621,377){\circle*{12}}
\put(343,348){\circle*{12}}
\put(245,342){\circle*{12}}
\put(621.0,373.0){\rule[-0.200pt]{0.400pt}{1.686pt}}
\put(611.0,373.0){\rule[-0.200pt]{4.818pt}{0.400pt}}
\put(611.0,380.0){\rule[-0.200pt]{4.818pt}{0.400pt}}
\put(343.0,345.0){\rule[-0.200pt]{0.400pt}{1.445pt}}
\put(333.0,345.0){\rule[-0.200pt]{4.818pt}{0.400pt}}
\put(333.0,351.0){\rule[-0.200pt]{4.818pt}{0.400pt}}
\put(245.0,337.0){\rule[-0.200pt]{0.400pt}{2.409pt}}
\put(235.0,337.0){\rule[-0.200pt]{4.818pt}{0.400pt}}
\put(235.0,347.0){\rule[-0.200pt]{4.818pt}{0.400pt}}
\put(621,530){\makebox(0,0){$\times$}}
\put(343,508){\makebox(0,0){$\times$}}
\put(245,501){\makebox(0,0){$\times$}}
\put(621.0,525.0){\rule[-0.200pt]{0.400pt}{2.650pt}}
\put(611.0,525.0){\rule[-0.200pt]{4.818pt}{0.400pt}}
\put(611.0,536.0){\rule[-0.200pt]{4.818pt}{0.400pt}}
\put(343.0,502.0){\rule[-0.200pt]{0.400pt}{3.132pt}}
\put(333.0,502.0){\rule[-0.200pt]{4.818pt}{0.400pt}}
\put(333.0,515.0){\rule[-0.200pt]{4.818pt}{0.400pt}}
\put(245.0,489.0){\rule[-0.200pt]{0.400pt}{6.022pt}}
\put(235.0,489.0){\rule[-0.200pt]{4.818pt}{0.400pt}}
\put(235.0,514.0){\rule[-0.200pt]{4.818pt}{0.400pt}}
\put(621,607){\makebox(0,0){$\star$}}
\put(343,553){\makebox(0,0){$\star$}}
\put(245,551){\makebox(0,0){$\star$}}
\put(621.0,597.0){\rule[-0.200pt]{0.400pt}{4.577pt}}
\put(611.0,597.0){\rule[-0.200pt]{4.818pt}{0.400pt}}
\put(611.0,616.0){\rule[-0.200pt]{4.818pt}{0.400pt}}
\put(343.0,546.0){\rule[-0.200pt]{0.400pt}{3.373pt}}
\put(333.0,546.0){\rule[-0.200pt]{4.818pt}{0.400pt}}
\put(333.0,560.0){\rule[-0.200pt]{4.818pt}{0.400pt}}
\put(245.0,541.0){\rule[-0.200pt]{0.400pt}{4.818pt}}
\put(235.0,541.0){\rule[-0.200pt]{4.818pt}{0.400pt}}
\put(235.0,561.0){\rule[-0.200pt]{4.818pt}{0.400pt}}
\put(621,764){\raisebox{-.8pt}{\makebox(0,0){$\Diamond$}}}
\put(343,708){\raisebox{-.8pt}{\makebox(0,0){$\Diamond$}}}
\put(245,727){\raisebox{-.8pt}{\makebox(0,0){$\Diamond$}}}
\put(621.0,741.0){\rule[-0.200pt]{0.400pt}{11.081pt}}
\put(611.0,741.0){\rule[-0.200pt]{4.818pt}{0.400pt}}
\put(611.0,787.0){\rule[-0.200pt]{4.818pt}{0.400pt}}
\put(343.0,686.0){\rule[-0.200pt]{0.400pt}{10.359pt}}
\put(333.0,686.0){\rule[-0.200pt]{4.818pt}{0.400pt}}
\put(333.0,729.0){\rule[-0.200pt]{4.818pt}{0.400pt}}
\put(245.0,700.0){\rule[-0.200pt]{0.400pt}{13.249pt}}
\put(235.0,700.0){\rule[-0.200pt]{4.818pt}{0.400pt}}
\put(235.0,755.0){\rule[-0.200pt]{4.818pt}{0.400pt}}
\sbox{\plotpoint}{\rule[-0.500pt]{1.000pt}{1.000pt}}%
\put(120,327){\usebox{\plotpoint}}
\put(120.00,327.00){\usebox{\plotpoint}}
\multiput(127,328)(20.473,3.412){0}{\usebox{\plotpoint}}
\multiput(133,329)(20.756,0.000){0}{\usebox{\plotpoint}}
\put(140.60,329.09){\usebox{\plotpoint}}
\multiput(147,330)(20.473,3.412){0}{\usebox{\plotpoint}}
\multiput(153,331)(20.756,0.000){0}{\usebox{\plotpoint}}
\put(161.19,331.17){\usebox{\plotpoint}}
\multiput(167,332)(20.756,0.000){0}{\usebox{\plotpoint}}
\multiput(173,332)(20.547,2.935){0}{\usebox{\plotpoint}}
\put(181.80,333.26){\usebox{\plotpoint}}
\multiput(187,334)(20.756,0.000){0}{\usebox{\plotpoint}}
\multiput(193,334)(20.547,2.935){0}{\usebox{\plotpoint}}
\put(202.41,335.34){\usebox{\plotpoint}}
\multiput(207,336)(20.756,0.000){0}{\usebox{\plotpoint}}
\multiput(213,336)(20.547,2.935){0}{\usebox{\plotpoint}}
\put(223.01,337.43){\usebox{\plotpoint}}
\multiput(227,338)(20.756,0.000){0}{\usebox{\plotpoint}}
\multiput(234,338)(20.473,3.412){0}{\usebox{\plotpoint}}
\put(243.61,339.52){\usebox{\plotpoint}}
\multiput(247,340)(20.756,0.000){0}{\usebox{\plotpoint}}
\multiput(254,340)(20.473,3.412){0}{\usebox{\plotpoint}}
\put(264.20,341.60){\usebox{\plotpoint}}
\multiput(267,342)(20.756,0.000){0}{\usebox{\plotpoint}}
\multiput(274,342)(20.473,3.412){0}{\usebox{\plotpoint}}
\put(284.80,343.69){\usebox{\plotpoint}}
\multiput(287,344)(20.756,0.000){0}{\usebox{\plotpoint}}
\multiput(294,344)(20.473,3.412){0}{\usebox{\plotpoint}}
\put(305.40,345.77){\usebox{\plotpoint}}
\multiput(307,346)(20.756,0.000){0}{\usebox{\plotpoint}}
\multiput(314,346)(20.473,3.412){0}{\usebox{\plotpoint}}
\put(326.05,347.00){\usebox{\plotpoint}}
\multiput(327,347)(20.547,2.935){0}{\usebox{\plotpoint}}
\multiput(334,348)(20.473,3.412){0}{\usebox{\plotpoint}}
\put(346.65,349.00){\usebox{\plotpoint}}
\multiput(347,349)(20.547,2.935){0}{\usebox{\plotpoint}}
\multiput(354,350)(20.473,3.412){0}{\usebox{\plotpoint}}
\multiput(360,351)(20.756,0.000){0}{\usebox{\plotpoint}}
\put(367.25,351.04){\usebox{\plotpoint}}
\multiput(374,352)(20.473,3.412){0}{\usebox{\plotpoint}}
\multiput(380,353)(20.756,0.000){0}{\usebox{\plotpoint}}
\put(387.85,353.12){\usebox{\plotpoint}}
\multiput(394,354)(20.473,3.412){0}{\usebox{\plotpoint}}
\multiput(400,355)(20.756,0.000){0}{\usebox{\plotpoint}}
\put(408.44,355.21){\usebox{\plotpoint}}
\multiput(414,356)(20.473,3.412){0}{\usebox{\plotpoint}}
\multiput(420,357)(20.756,0.000){0}{\usebox{\plotpoint}}
\put(429.04,357.29){\usebox{\plotpoint}}
\multiput(434,358)(20.473,3.412){0}{\usebox{\plotpoint}}
\multiput(440,359)(20.756,0.000){0}{\usebox{\plotpoint}}
\put(449.64,359.38){\usebox{\plotpoint}}
\multiput(454,360)(20.547,2.935){0}{\usebox{\plotpoint}}
\multiput(461,361)(20.756,0.000){0}{\usebox{\plotpoint}}
\put(470.24,361.46){\usebox{\plotpoint}}
\multiput(474,362)(20.756,0.000){0}{\usebox{\plotpoint}}
\multiput(481,362)(20.473,3.412){0}{\usebox{\plotpoint}}
\put(490.84,363.55){\usebox{\plotpoint}}
\multiput(494,364)(20.756,0.000){0}{\usebox{\plotpoint}}
\multiput(501,364)(20.473,3.412){0}{\usebox{\plotpoint}}
\put(511.43,365.63){\usebox{\plotpoint}}
\multiput(514,366)(20.756,0.000){0}{\usebox{\plotpoint}}
\multiput(521,366)(20.473,3.412){0}{\usebox{\plotpoint}}
\put(532.03,367.72){\usebox{\plotpoint}}
\multiput(534,368)(20.756,0.000){0}{\usebox{\plotpoint}}
\multiput(541,368)(20.473,3.412){0}{\usebox{\plotpoint}}
\put(552.62,369.80){\usebox{\plotpoint}}
\multiput(554,370)(20.756,0.000){0}{\usebox{\plotpoint}}
\multiput(561,370)(20.473,3.412){0}{\usebox{\plotpoint}}
\put(573.22,371.89){\usebox{\plotpoint}}
\multiput(574,372)(20.756,0.000){0}{\usebox{\plotpoint}}
\multiput(581,372)(20.473,3.412){0}{\usebox{\plotpoint}}
\put(593.82,373.97){\usebox{\plotpoint}}
\multiput(594,374)(20.756,0.000){0}{\usebox{\plotpoint}}
\multiput(601,374)(20.473,3.412){0}{\usebox{\plotpoint}}
\multiput(607,375)(20.547,2.935){0}{\usebox{\plotpoint}}
\put(614.42,376.00){\usebox{\plotpoint}}
\multiput(621,376)(20.473,3.412){0}{\usebox{\plotpoint}}
\multiput(627,377)(20.756,0.000){0}{\usebox{\plotpoint}}
\put(635.08,377.15){\usebox{\plotpoint}}
\multiput(641,378)(20.473,3.412){0}{\usebox{\plotpoint}}
\multiput(647,379)(20.756,0.000){0}{\usebox{\plotpoint}}
\put(655.67,379.24){\usebox{\plotpoint}}
\multiput(661,380)(20.473,3.412){0}{\usebox{\plotpoint}}
\multiput(667,381)(20.756,0.000){0}{\usebox{\plotpoint}}
\put(676.27,381.32){\usebox{\plotpoint}}
\multiput(681,382)(20.547,2.935){0}{\usebox{\plotpoint}}
\multiput(688,383)(20.756,0.000){0}{\usebox{\plotpoint}}
\put(696.88,383.41){\usebox{\plotpoint}}
\multiput(701,384)(20.547,2.935){0}{\usebox{\plotpoint}}
\multiput(708,385)(20.756,0.000){0}{\usebox{\plotpoint}}
\put(717.48,385.50){\usebox{\plotpoint}}
\multiput(721,386)(20.547,2.935){0}{\usebox{\plotpoint}}
\multiput(728,387)(20.756,0.000){0}{\usebox{\plotpoint}}
\put(738.09,387.58){\usebox{\plotpoint}}
\multiput(741,388)(20.547,2.935){0}{\usebox{\plotpoint}}
\multiput(748,389)(20.756,0.000){0}{\usebox{\plotpoint}}
\put(758.70,389.67){\usebox{\plotpoint}}
\multiput(761,390)(20.547,2.935){0}{\usebox{\plotpoint}}
\multiput(768,391)(20.756,0.000){0}{\usebox{\plotpoint}}
\put(779.30,391.76){\usebox{\plotpoint}}
\put(781,392){\usebox{\plotpoint}}
\put(120,491){\usebox{\plotpoint}}
\put(120.00,491.00){\usebox{\plotpoint}}
\multiput(127,491)(20.473,3.412){0}{\usebox{\plotpoint}}
\multiput(133,492)(20.547,2.935){0}{\usebox{\plotpoint}}
\put(140.60,493.00){\usebox{\plotpoint}}
\multiput(147,493)(20.473,3.412){0}{\usebox{\plotpoint}}
\multiput(153,494)(20.756,0.000){0}{\usebox{\plotpoint}}
\put(161.26,494.18){\usebox{\plotpoint}}
\multiput(167,495)(20.756,0.000){0}{\usebox{\plotpoint}}
\multiput(173,495)(20.547,2.935){0}{\usebox{\plotpoint}}
\put(181.89,496.00){\usebox{\plotpoint}}
\multiput(187,496)(20.473,3.412){0}{\usebox{\plotpoint}}
\multiput(193,497)(20.756,0.000){0}{\usebox{\plotpoint}}
\put(202.53,497.36){\usebox{\plotpoint}}
\multiput(207,498)(20.756,0.000){0}{\usebox{\plotpoint}}
\multiput(213,498)(20.547,2.935){0}{\usebox{\plotpoint}}
\put(223.17,499.00){\usebox{\plotpoint}}
\multiput(227,499)(20.547,2.935){0}{\usebox{\plotpoint}}
\multiput(234,500)(20.756,0.000){0}{\usebox{\plotpoint}}
\put(243.82,500.55){\usebox{\plotpoint}}
\multiput(247,501)(20.547,2.935){0}{\usebox{\plotpoint}}
\multiput(254,502)(20.756,0.000){0}{\usebox{\plotpoint}}
\put(264.43,502.63){\usebox{\plotpoint}}
\multiput(267,503)(20.756,0.000){0}{\usebox{\plotpoint}}
\multiput(274,503)(20.473,3.412){0}{\usebox{\plotpoint}}
\put(285.07,504.00){\usebox{\plotpoint}}
\multiput(287,504)(20.547,2.935){0}{\usebox{\plotpoint}}
\multiput(294,505)(20.756,0.000){0}{\usebox{\plotpoint}}
\put(305.70,505.81){\usebox{\plotpoint}}
\multiput(307,506)(20.756,0.000){0}{\usebox{\plotpoint}}
\multiput(314,506)(20.473,3.412){0}{\usebox{\plotpoint}}
\put(326.36,507.00){\usebox{\plotpoint}}
\multiput(327,507)(20.547,2.935){0}{\usebox{\plotpoint}}
\multiput(334,508)(20.756,0.000){0}{\usebox{\plotpoint}}
\put(346.97,509.00){\usebox{\plotpoint}}
\multiput(347,509)(20.756,0.000){0}{\usebox{\plotpoint}}
\multiput(354,509)(20.473,3.412){0}{\usebox{\plotpoint}}
\multiput(360,510)(20.756,0.000){0}{\usebox{\plotpoint}}
\put(367.64,510.09){\usebox{\plotpoint}}
\multiput(374,511)(20.473,3.412){0}{\usebox{\plotpoint}}
\multiput(380,512)(20.756,0.000){0}{\usebox{\plotpoint}}
\put(388.23,512.18){\usebox{\plotpoint}}
\multiput(394,513)(20.756,0.000){0}{\usebox{\plotpoint}}
\multiput(400,513)(20.547,2.935){0}{\usebox{\plotpoint}}
\put(408.86,514.00){\usebox{\plotpoint}}
\multiput(414,514)(20.473,3.412){0}{\usebox{\plotpoint}}
\multiput(420,515)(20.756,0.000){0}{\usebox{\plotpoint}}
\put(429.51,515.36){\usebox{\plotpoint}}
\multiput(434,516)(20.756,0.000){0}{\usebox{\plotpoint}}
\multiput(440,516)(20.547,2.935){0}{\usebox{\plotpoint}}
\put(450.15,517.00){\usebox{\plotpoint}}
\multiput(454,517)(20.547,2.935){0}{\usebox{\plotpoint}}
\multiput(461,518)(20.756,0.000){0}{\usebox{\plotpoint}}
\put(470.79,518.54){\usebox{\plotpoint}}
\multiput(474,519)(20.756,0.000){0}{\usebox{\plotpoint}}
\multiput(481,519)(20.473,3.412){0}{\usebox{\plotpoint}}
\put(491.39,520.63){\usebox{\plotpoint}}
\multiput(494,521)(20.756,0.000){0}{\usebox{\plotpoint}}
\multiput(501,521)(20.473,3.412){0}{\usebox{\plotpoint}}
\put(512.03,522.00){\usebox{\plotpoint}}
\multiput(514,522)(20.547,2.935){0}{\usebox{\plotpoint}}
\multiput(521,523)(20.756,0.000){0}{\usebox{\plotpoint}}
\put(532.66,523.81){\usebox{\plotpoint}}
\multiput(534,524)(20.756,0.000){0}{\usebox{\plotpoint}}
\multiput(541,524)(20.473,3.412){0}{\usebox{\plotpoint}}
\put(553.32,525.00){\usebox{\plotpoint}}
\multiput(554,525)(20.547,2.935){0}{\usebox{\plotpoint}}
\multiput(561,526)(20.756,0.000){0}{\usebox{\plotpoint}}
\put(573.93,526.99){\usebox{\plotpoint}}
\multiput(574,527)(20.756,0.000){0}{\usebox{\plotpoint}}
\multiput(581,527)(20.473,3.412){0}{\usebox{\plotpoint}}
\multiput(587,528)(20.756,0.000){0}{\usebox{\plotpoint}}
\put(594.60,528.09){\usebox{\plotpoint}}
\multiput(601,529)(20.473,3.412){0}{\usebox{\plotpoint}}
\multiput(607,530)(20.756,0.000){0}{\usebox{\plotpoint}}
\put(615.20,530.17){\usebox{\plotpoint}}
\multiput(621,531)(20.756,0.000){0}{\usebox{\plotpoint}}
\multiput(627,531)(20.547,2.935){0}{\usebox{\plotpoint}}
\put(635.82,532.00){\usebox{\plotpoint}}
\multiput(641,532)(20.473,3.412){0}{\usebox{\plotpoint}}
\multiput(647,533)(20.756,0.000){0}{\usebox{\plotpoint}}
\put(656.47,533.35){\usebox{\plotpoint}}
\multiput(661,534)(20.756,0.000){0}{\usebox{\plotpoint}}
\multiput(667,534)(20.547,2.935){0}{\usebox{\plotpoint}}
\put(677.11,535.00){\usebox{\plotpoint}}
\multiput(681,535)(20.547,2.935){0}{\usebox{\plotpoint}}
\multiput(688,536)(20.756,0.000){0}{\usebox{\plotpoint}}
\put(697.75,536.54){\usebox{\plotpoint}}
\multiput(701,537)(20.756,0.000){0}{\usebox{\plotpoint}}
\multiput(708,537)(20.473,3.412){0}{\usebox{\plotpoint}}
\put(718.39,538.00){\usebox{\plotpoint}}
\multiput(721,538)(20.547,2.935){0}{\usebox{\plotpoint}}
\multiput(728,539)(20.473,3.412){0}{\usebox{\plotpoint}}
\put(739.00,540.00){\usebox{\plotpoint}}
\multiput(741,540)(20.547,2.935){0}{\usebox{\plotpoint}}
\multiput(748,541)(20.756,0.000){0}{\usebox{\plotpoint}}
\put(759.62,541.80){\usebox{\plotpoint}}
\multiput(761,542)(20.756,0.000){0}{\usebox{\plotpoint}}
\multiput(768,542)(20.473,3.412){0}{\usebox{\plotpoint}}
\put(780.28,543.00){\usebox{\plotpoint}}
\put(781,543){\usebox{\plotpoint}}
\put(120,522){\usebox{\plotpoint}}
\put(120.00,522.00){\usebox{\plotpoint}}
\multiput(127,523)(19.690,6.563){0}{\usebox{\plotpoint}}
\multiput(133,525)(20.547,2.935){0}{\usebox{\plotpoint}}
\put(140.29,526.04){\usebox{\plotpoint}}
\multiput(147,527)(20.473,3.412){0}{\usebox{\plotpoint}}
\multiput(153,528)(20.547,2.935){0}{\usebox{\plotpoint}}
\put(160.81,529.12){\usebox{\plotpoint}}
\multiput(167,530)(20.473,3.412){0}{\usebox{\plotpoint}}
\multiput(173,531)(20.547,2.935){0}{\usebox{\plotpoint}}
\put(181.34,532.19){\usebox{\plotpoint}}
\multiput(187,533)(20.473,3.412){0}{\usebox{\plotpoint}}
\multiput(193,534)(20.547,2.935){0}{\usebox{\plotpoint}}
\put(201.81,535.52){\usebox{\plotpoint}}
\multiput(207,537)(20.473,3.412){0}{\usebox{\plotpoint}}
\multiput(213,538)(20.547,2.935){0}{\usebox{\plotpoint}}
\put(222.18,539.31){\usebox{\plotpoint}}
\multiput(227,540)(20.547,2.935){0}{\usebox{\plotpoint}}
\multiput(234,541)(20.473,3.412){0}{\usebox{\plotpoint}}
\put(242.71,542.39){\usebox{\plotpoint}}
\multiput(247,543)(20.547,2.935){0}{\usebox{\plotpoint}}
\multiput(254,544)(20.473,3.412){0}{\usebox{\plotpoint}}
\put(263.23,545.46){\usebox{\plotpoint}}
\multiput(267,546)(20.547,2.935){0}{\usebox{\plotpoint}}
\multiput(274,547)(19.690,6.563){0}{\usebox{\plotpoint}}
\put(283.52,549.50){\usebox{\plotpoint}}
\multiput(287,550)(20.547,2.935){0}{\usebox{\plotpoint}}
\multiput(294,551)(20.473,3.412){0}{\usebox{\plotpoint}}
\put(304.04,552.58){\usebox{\plotpoint}}
\multiput(307,553)(20.547,2.935){0}{\usebox{\plotpoint}}
\multiput(314,554)(20.473,3.412){0}{\usebox{\plotpoint}}
\put(324.57,555.65){\usebox{\plotpoint}}
\multiput(327,556)(20.547,2.935){0}{\usebox{\plotpoint}}
\multiput(334,557)(20.473,3.412){0}{\usebox{\plotpoint}}
\put(345.09,558.73){\usebox{\plotpoint}}
\multiput(347,559)(19.957,5.702){0}{\usebox{\plotpoint}}
\multiput(354,561)(20.473,3.412){0}{\usebox{\plotpoint}}
\put(365.41,562.77){\usebox{\plotpoint}}
\multiput(367,563)(20.547,2.935){0}{\usebox{\plotpoint}}
\multiput(374,564)(20.473,3.412){0}{\usebox{\plotpoint}}
\put(385.94,565.85){\usebox{\plotpoint}}
\multiput(387,566)(20.547,2.935){0}{\usebox{\plotpoint}}
\multiput(394,567)(20.473,3.412){0}{\usebox{\plotpoint}}
\put(406.46,568.92){\usebox{\plotpoint}}
\multiput(407,569)(20.547,2.935){0}{\usebox{\plotpoint}}
\multiput(414,570)(20.473,3.412){0}{\usebox{\plotpoint}}
\put(426.79,572.94){\usebox{\plotpoint}}
\multiput(427,573)(20.547,2.935){0}{\usebox{\plotpoint}}
\multiput(434,574)(20.473,3.412){0}{\usebox{\plotpoint}}
\multiput(440,575)(20.547,2.935){0}{\usebox{\plotpoint}}
\put(447.30,576.04){\usebox{\plotpoint}}
\multiput(454,577)(20.547,2.935){0}{\usebox{\plotpoint}}
\multiput(461,578)(20.473,3.412){0}{\usebox{\plotpoint}}
\put(467.83,579.12){\usebox{\plotpoint}}
\multiput(474,580)(20.547,2.935){0}{\usebox{\plotpoint}}
\multiput(481,581)(20.473,3.412){0}{\usebox{\plotpoint}}
\put(488.36,582.19){\usebox{\plotpoint}}
\multiput(494,583)(19.957,5.702){0}{\usebox{\plotpoint}}
\multiput(501,585)(20.473,3.412){0}{\usebox{\plotpoint}}
\put(508.67,586.24){\usebox{\plotpoint}}
\multiput(514,587)(20.547,2.935){0}{\usebox{\plotpoint}}
\multiput(521,588)(20.473,3.412){0}{\usebox{\plotpoint}}
\put(529.20,589.31){\usebox{\plotpoint}}
\multiput(534,590)(20.547,2.935){0}{\usebox{\plotpoint}}
\multiput(541,591)(20.473,3.412){0}{\usebox{\plotpoint}}
\put(549.72,592.39){\usebox{\plotpoint}}
\multiput(554,593)(20.547,2.935){0}{\usebox{\plotpoint}}
\multiput(561,594)(20.473,3.412){0}{\usebox{\plotpoint}}
\put(570.16,595.90){\usebox{\plotpoint}}
\multiput(574,597)(20.547,2.935){0}{\usebox{\plotpoint}}
\multiput(581,598)(20.473,3.412){0}{\usebox{\plotpoint}}
\put(590.57,599.51){\usebox{\plotpoint}}
\multiput(594,600)(20.547,2.935){0}{\usebox{\plotpoint}}
\multiput(601,601)(20.473,3.412){0}{\usebox{\plotpoint}}
\put(611.09,602.58){\usebox{\plotpoint}}
\multiput(614,603)(20.547,2.935){0}{\usebox{\plotpoint}}
\multiput(621,604)(20.473,3.412){0}{\usebox{\plotpoint}}
\put(631.62,605.66){\usebox{\plotpoint}}
\multiput(634,606)(20.547,2.935){0}{\usebox{\plotpoint}}
\multiput(641,607)(19.690,6.563){0}{\usebox{\plotpoint}}
\put(651.90,609.70){\usebox{\plotpoint}}
\multiput(654,610)(20.547,2.935){0}{\usebox{\plotpoint}}
\multiput(661,611)(20.473,3.412){0}{\usebox{\plotpoint}}
\put(672.43,612.78){\usebox{\plotpoint}}
\multiput(674,613)(20.547,2.935){0}{\usebox{\plotpoint}}
\multiput(681,614)(20.547,2.935){0}{\usebox{\plotpoint}}
\put(692.96,615.83){\usebox{\plotpoint}}
\multiput(694,616)(20.547,2.935){0}{\usebox{\plotpoint}}
\multiput(701,617)(20.547,2.935){0}{\usebox{\plotpoint}}
\put(713.48,618.91){\usebox{\plotpoint}}
\multiput(714,619)(20.547,2.935){0}{\usebox{\plotpoint}}
\multiput(721,620)(19.957,5.702){0}{\usebox{\plotpoint}}
\put(733.80,622.97){\usebox{\plotpoint}}
\multiput(734,623)(20.547,2.935){0}{\usebox{\plotpoint}}
\multiput(741,624)(20.547,2.935){0}{\usebox{\plotpoint}}
\multiput(748,625)(20.473,3.412){0}{\usebox{\plotpoint}}
\put(754.32,626.05){\usebox{\plotpoint}}
\multiput(761,627)(20.547,2.935){0}{\usebox{\plotpoint}}
\multiput(768,628)(20.473,3.412){0}{\usebox{\plotpoint}}
\put(774.85,629.12){\usebox{\plotpoint}}
\put(781,630){\usebox{\plotpoint}}
\put(120,693){\usebox{\plotpoint}}
\put(120.00,693.00){\usebox{\plotpoint}}
\multiput(127,694)(20.473,3.412){0}{\usebox{\plotpoint}}
\multiput(133,695)(20.756,0.000){0}{\usebox{\plotpoint}}
\put(140.60,695.09){\usebox{\plotpoint}}
\multiput(147,696)(20.473,3.412){0}{\usebox{\plotpoint}}
\multiput(153,697)(20.547,2.935){0}{\usebox{\plotpoint}}
\put(161.12,698.16){\usebox{\plotpoint}}
\multiput(167,699)(20.473,3.412){0}{\usebox{\plotpoint}}
\multiput(173,700)(20.547,2.935){0}{\usebox{\plotpoint}}
\put(181.65,701.24){\usebox{\plotpoint}}
\multiput(187,702)(20.473,3.412){0}{\usebox{\plotpoint}}
\multiput(193,703)(20.756,0.000){0}{\usebox{\plotpoint}}
\put(202.24,703.32){\usebox{\plotpoint}}
\multiput(207,704)(20.473,3.412){0}{\usebox{\plotpoint}}
\multiput(213,705)(20.547,2.935){0}{\usebox{\plotpoint}}
\put(222.77,706.40){\usebox{\plotpoint}}
\multiput(227,707)(20.547,2.935){0}{\usebox{\plotpoint}}
\multiput(234,708)(20.473,3.412){0}{\usebox{\plotpoint}}
\put(243.29,709.47){\usebox{\plotpoint}}
\multiput(247,710)(20.547,2.935){0}{\usebox{\plotpoint}}
\multiput(254,711)(20.473,3.412){0}{\usebox{\plotpoint}}
\put(263.86,712.00){\usebox{\plotpoint}}
\multiput(267,712)(20.547,2.935){0}{\usebox{\plotpoint}}
\multiput(274,713)(20.473,3.412){0}{\usebox{\plotpoint}}
\put(284.41,714.63){\usebox{\plotpoint}}
\multiput(287,715)(20.547,2.935){0}{\usebox{\plotpoint}}
\multiput(294,716)(20.473,3.412){0}{\usebox{\plotpoint}}
\put(304.94,717.71){\usebox{\plotpoint}}
\multiput(307,718)(20.547,2.935){0}{\usebox{\plotpoint}}
\multiput(314,719)(20.473,3.412){0}{\usebox{\plotpoint}}
\put(325.52,720.00){\usebox{\plotpoint}}
\multiput(327,720)(20.547,2.935){0}{\usebox{\plotpoint}}
\multiput(334,721)(20.473,3.412){0}{\usebox{\plotpoint}}
\put(346.06,722.87){\usebox{\plotpoint}}
\multiput(347,723)(20.547,2.935){0}{\usebox{\plotpoint}}
\multiput(354,724)(20.473,3.412){0}{\usebox{\plotpoint}}
\put(366.58,725.94){\usebox{\plotpoint}}
\multiput(367,726)(20.547,2.935){0}{\usebox{\plotpoint}}
\multiput(374,727)(20.473,3.412){0}{\usebox{\plotpoint}}
\multiput(380,728)(20.756,0.000){0}{\usebox{\plotpoint}}
\put(387.18,728.03){\usebox{\plotpoint}}
\multiput(394,729)(20.473,3.412){0}{\usebox{\plotpoint}}
\multiput(400,730)(20.547,2.935){0}{\usebox{\plotpoint}}
\put(407.71,731.10){\usebox{\plotpoint}}
\multiput(414,732)(20.473,3.412){0}{\usebox{\plotpoint}}
\multiput(420,733)(20.547,2.935){0}{\usebox{\plotpoint}}
\put(428.23,734.18){\usebox{\plotpoint}}
\multiput(434,735)(20.473,3.412){0}{\usebox{\plotpoint}}
\multiput(440,736)(20.756,0.000){0}{\usebox{\plotpoint}}
\put(448.83,736.26){\usebox{\plotpoint}}
\multiput(454,737)(20.547,2.935){0}{\usebox{\plotpoint}}
\multiput(461,738)(20.473,3.412){0}{\usebox{\plotpoint}}
\put(469.35,739.34){\usebox{\plotpoint}}
\multiput(474,740)(20.547,2.935){0}{\usebox{\plotpoint}}
\multiput(481,741)(20.473,3.412){0}{\usebox{\plotpoint}}
\put(489.88,742.41){\usebox{\plotpoint}}
\multiput(494,743)(20.547,2.935){0}{\usebox{\plotpoint}}
\multiput(501,744)(20.756,0.000){0}{\usebox{\plotpoint}}
\put(510.48,744.50){\usebox{\plotpoint}}
\multiput(514,745)(20.547,2.935){0}{\usebox{\plotpoint}}
\multiput(521,746)(20.473,3.412){0}{\usebox{\plotpoint}}
\put(531.01,747.57){\usebox{\plotpoint}}
\multiput(534,748)(20.547,2.935){0}{\usebox{\plotpoint}}
\multiput(541,749)(20.473,3.412){0}{\usebox{\plotpoint}}
\put(551.53,750.65){\usebox{\plotpoint}}
\multiput(554,751)(20.547,2.935){0}{\usebox{\plotpoint}}
\multiput(561,752)(20.756,0.000){0}{\usebox{\plotpoint}}
\put(572.14,752.73){\usebox{\plotpoint}}
\multiput(574,753)(20.547,2.935){0}{\usebox{\plotpoint}}
\multiput(581,754)(20.473,3.412){0}{\usebox{\plotpoint}}
\put(592.67,755.81){\usebox{\plotpoint}}
\multiput(594,756)(20.547,2.935){0}{\usebox{\plotpoint}}
\multiput(601,757)(20.473,3.412){0}{\usebox{\plotpoint}}
\put(613.19,758.88){\usebox{\plotpoint}}
\multiput(614,759)(20.547,2.935){0}{\usebox{\plotpoint}}
\multiput(621,760)(20.473,3.412){0}{\usebox{\plotpoint}}
\put(633.79,761.00){\usebox{\plotpoint}}
\multiput(634,761)(20.547,2.935){0}{\usebox{\plotpoint}}
\multiput(641,762)(20.473,3.412){0}{\usebox{\plotpoint}}
\multiput(647,763)(20.547,2.935){0}{\usebox{\plotpoint}}
\put(654.31,764.04){\usebox{\plotpoint}}
\multiput(661,765)(20.473,3.412){0}{\usebox{\plotpoint}}
\multiput(667,766)(20.547,2.935){0}{\usebox{\plotpoint}}
\put(674.84,767.12){\usebox{\plotpoint}}
\multiput(681,768)(20.547,2.935){0}{\usebox{\plotpoint}}
\multiput(688,769)(20.756,0.000){0}{\usebox{\plotpoint}}
\put(695.45,769.21){\usebox{\plotpoint}}
\multiput(701,770)(20.547,2.935){0}{\usebox{\plotpoint}}
\multiput(708,771)(20.473,3.412){0}{\usebox{\plotpoint}}
\put(715.97,772.28){\usebox{\plotpoint}}
\multiput(721,773)(20.547,2.935){0}{\usebox{\plotpoint}}
\multiput(728,774)(20.473,3.412){0}{\usebox{\plotpoint}}
\put(736.50,775.36){\usebox{\plotpoint}}
\multiput(741,776)(20.547,2.935){0}{\usebox{\plotpoint}}
\multiput(748,777)(20.756,0.000){0}{\usebox{\plotpoint}}
\put(757.10,777.44){\usebox{\plotpoint}}
\multiput(761,778)(20.547,2.935){0}{\usebox{\plotpoint}}
\multiput(768,779)(20.473,3.412){0}{\usebox{\plotpoint}}
\put(777.63,780.52){\usebox{\plotpoint}}
\put(781,781){\usebox{\plotpoint}}
\end{picture}

%% file: plot_glue4.tex
% GNUPLOT: LaTeX picture
\setlength{\unitlength}{0.240900pt}
\ifx\plotpoint\undefined\newsavebox{\plotpoint}\fi
\sbox{\plotpoint}{\rule[-0.200pt]{0.400pt}{0.400pt}}%
\begin{picture}(825,675)(0,0)
\font\gnuplot=cmr10 at 12pt
\gnuplot
\sbox{\plotpoint}{\rule[-0.200pt]{0.400pt}{0.400pt}}%
\put(120.0,31.0){\rule[-0.200pt]{4.818pt}{0.400pt}}
\put(108,31){\makebox(0,0)[r]{{$0$}}}
\put(761.0,31.0){\rule[-0.200pt]{4.818pt}{0.400pt}}
\put(120.0,190.0){\rule[-0.200pt]{4.818pt}{0.400pt}}
\put(108,190){\makebox(0,0)[r]{{$2$}}}
\put(761.0,190.0){\rule[-0.200pt]{4.818pt}{0.400pt}}
\put(120.0,350.0){\rule[-0.200pt]{4.818pt}{0.400pt}}
\put(108,350){\makebox(0,0)[r]{{$4$}}}
\put(761.0,350.0){\rule[-0.200pt]{4.818pt}{0.400pt}}
\put(120.0,509.0){\rule[-0.200pt]{4.818pt}{0.400pt}}
\put(108,509){\makebox(0,0)[r]{{$6$}}}
\put(761.0,509.0){\rule[-0.200pt]{4.818pt}{0.400pt}}
\put(320.0,31.0){\rule[-0.200pt]{0.400pt}{4.818pt}}
\put(320,19){\makebox(0,0){\shortstack{\\ \\ \\ {$0.1$}}}}
\put(320.0,648.0){\rule[-0.200pt]{0.400pt}{4.818pt}}
\put(521.0,31.0){\rule[-0.200pt]{0.400pt}{4.818pt}}
\put(521,19){\makebox(0,0){\shortstack{\\ \\ \\ {$0.2$}}}}
\put(521.0,648.0){\rule[-0.200pt]{0.400pt}{4.818pt}}
\put(721.0,31.0){\rule[-0.200pt]{0.400pt}{4.818pt}}
\put(721,19){\makebox(0,0){\shortstack{\\ \\ \\ {$0.3$}}}}
\put(721.0,648.0){\rule[-0.200pt]{0.400pt}{4.818pt}}
\put(120.0,31.0){\rule[-0.200pt]{159.235pt}{0.400pt}}
\put(781.0,31.0){\rule[-0.200pt]{0.400pt}{153.453pt}}
\put(120.0,668.0){\rule[-0.200pt]{159.235pt}{0.400pt}}
\put(12,349){\makebox(0,0){{${m \over {\surd\sigma}}$}}}
\put(450,-53){\makebox(0,0){{$1/N^2_c$}}}
\put(120.0,31.0){\rule[-0.200pt]{0.400pt}{153.453pt}}
\put(621,339){\circle*{12}}
\put(343,321){\circle*{12}}
\put(248,300){\circle*{12}}
\put(242,348){\circle*{12}}
\put(621.0,330.0){\rule[-0.200pt]{0.400pt}{4.577pt}}
\put(611.0,330.0){\rule[-0.200pt]{4.818pt}{0.400pt}}
\put(611.0,349.0){\rule[-0.200pt]{4.818pt}{0.400pt}}
\put(343.0,309.0){\rule[-0.200pt]{0.400pt}{5.782pt}}
\put(333.0,309.0){\rule[-0.200pt]{4.818pt}{0.400pt}}
\put(333.0,333.0){\rule[-0.200pt]{4.818pt}{0.400pt}}
\put(248.0,275.0){\rule[-0.200pt]{0.400pt}{12.045pt}}
\put(238.0,275.0){\rule[-0.200pt]{4.818pt}{0.400pt}}
\put(238.0,325.0){\rule[-0.200pt]{4.818pt}{0.400pt}}
\put(242.0,324.0){\rule[-0.200pt]{0.400pt}{11.563pt}}
\put(232.0,324.0){\rule[-0.200pt]{4.818pt}{0.400pt}}
\put(232.0,372.0){\rule[-0.200pt]{4.818pt}{0.400pt}}
\put(621,479){\circle{12}}
\put(343,440){\circle{12}}
\put(248,431){\circle{12}}
\put(242,463){\circle{12}}
\put(621.0,471.0){\rule[-0.200pt]{0.400pt}{4.095pt}}
\put(611.0,471.0){\rule[-0.200pt]{4.818pt}{0.400pt}}
\put(611.0,488.0){\rule[-0.200pt]{4.818pt}{0.400pt}}
\put(343.0,422.0){\rule[-0.200pt]{0.400pt}{8.913pt}}
\put(333.0,422.0){\rule[-0.200pt]{4.818pt}{0.400pt}}
\put(333.0,459.0){\rule[-0.200pt]{4.818pt}{0.400pt}}
\put(248.0,398.0){\rule[-0.200pt]{0.400pt}{15.658pt}}
\put(238.0,398.0){\rule[-0.200pt]{4.818pt}{0.400pt}}
\put(238.0,463.0){\rule[-0.200pt]{4.818pt}{0.400pt}}
\put(242.0,423.0){\rule[-0.200pt]{0.400pt}{19.513pt}}
\put(232.0,423.0){\rule[-0.200pt]{4.818pt}{0.400pt}}
\put(232.0,504.0){\rule[-0.200pt]{4.818pt}{0.400pt}}
\end{picture}

%% file: plot_top4.tex
% GNUPLOT: LaTeX picture
\setlength{\unitlength}{0.240900pt}
\ifx\plotpoint\undefined\newsavebox{\plotpoint}\fi
\sbox{\plotpoint}{\rule[-0.200pt]{0.400pt}{0.400pt}}%
\begin{picture}(825,584)(0,0)
\font\gnuplot=cmr10 at 12pt
\gnuplot
\sbox{\plotpoint}{\rule[-0.200pt]{0.400pt}{0.400pt}}%
\put(120.0,31.0){\rule[-0.200pt]{4.818pt}{0.400pt}}
\put(108,31){\makebox(0,0)[r]{{$0$}}}
\put(761.0,31.0){\rule[-0.200pt]{4.818pt}{0.400pt}}
\put(120.0,177.0){\rule[-0.200pt]{4.818pt}{0.400pt}}
\put(108,177){\makebox(0,0)[r]{{$0.2$}}}
\put(761.0,177.0){\rule[-0.200pt]{4.818pt}{0.400pt}}
\put(120.0,322.0){\rule[-0.200pt]{4.818pt}{0.400pt}}
\put(108,322){\makebox(0,0)[r]{{$0.4$}}}
\put(761.0,322.0){\rule[-0.200pt]{4.818pt}{0.400pt}}
\put(120.0,468.0){\rule[-0.200pt]{4.818pt}{0.400pt}}
\put(108,468){\makebox(0,0)[r]{{$0.6$}}}
\put(761.0,468.0){\rule[-0.200pt]{4.818pt}{0.400pt}}
\put(320.0,31.0){\rule[-0.200pt]{0.400pt}{4.818pt}}
\put(320,19){\makebox(0,0){\shortstack{\\ \\ \\ {$0.1$}}}}
\put(320.0,557.0){\rule[-0.200pt]{0.400pt}{4.818pt}}
\put(521.0,31.0){\rule[-0.200pt]{0.400pt}{4.818pt}}
\put(521,19){\makebox(0,0){\shortstack{\\ \\ \\ {$0.2$}}}}
\put(521.0,557.0){\rule[-0.200pt]{0.400pt}{4.818pt}}
\put(721.0,31.0){\rule[-0.200pt]{0.400pt}{4.818pt}}
\put(721,19){\makebox(0,0){\shortstack{\\ \\ \\ {$0.3$}}}}
\put(721.0,557.0){\rule[-0.200pt]{0.400pt}{4.818pt}}
\put(120.0,31.0){\rule[-0.200pt]{159.235pt}{0.400pt}}
\put(781.0,31.0){\rule[-0.200pt]{0.400pt}{131.531pt}}
\put(120.0,577.0){\rule[-0.200pt]{159.235pt}{0.400pt}}
\put(12,400){\makebox(0,0){{${{\chi_t^{1\over 4}} \over {\surd\sigma}}$}}}
\put(450,-53){\makebox(0,0){{$1/N^2_c$}}}
\put(120.0,31.0){\rule[-0.200pt]{0.400pt}{131.531pt}}
\put(621,373){\circle*{12}}
\put(343,349){\circle*{12}}
\put(245,307){\circle*{12}}
\put(621.0,359.0){\rule[-0.200pt]{0.400pt}{6.745pt}}
\put(611.0,359.0){\rule[-0.200pt]{4.818pt}{0.400pt}}
\put(611.0,387.0){\rule[-0.200pt]{4.818pt}{0.400pt}}
\put(343.0,331.0){\rule[-0.200pt]{0.400pt}{8.672pt}}
\put(333.0,331.0){\rule[-0.200pt]{4.818pt}{0.400pt}}
\put(333.0,367.0){\rule[-0.200pt]{4.818pt}{0.400pt}}
\put(245.0,289.0){\rule[-0.200pt]{0.400pt}{8.672pt}}
\put(235.0,289.0){\rule[-0.200pt]{4.818pt}{0.400pt}}
\put(235.0,325.0){\rule[-0.200pt]{4.818pt}{0.400pt}}
\end{picture}